\documentclass[useAMS]{mn2e}
\usepackage{psfig,epsfig}
\usepackage{graphicx}
\usepackage{amssymb}
\usepackage{epstopdf}

\begin{document}
\def\simlt{\mathrel{\rlap{\lower 3pt\hbox{$\sim$}}
        \raise 2.0pt\hbox{$<$}}}
\def\simgt{\mathrel{\rlap{\lower 3pt\hbox{$\sim$}}
        \raise 2.0pt\hbox{$>$}}}

\title[Clustering of radio AGN and star-forming galaxies]{The clustering properties of radio-selected AGN and star-forming galaxies up to redshifts $z\sim 3$}

\author[Manuela Magliocchetti et al.]
{\parbox[t]\textwidth{M. Magliocchetti$^{1}$,  P. Popesso$^{2}$, M. Brusa$^{3,4}$, M. Salvato$^{5}$, C. Laigle$^{6}$, H.J. McCracken$^{6}$, O.Ilbert$^{7}$\\
} \\
{\tt $^1$ INAF-IAPS, Via Fosso del Cavaliere 100, 00133 Roma,
  Italy}\\
 {\tt $^2$ Excellence Cluster, Boltzmannstr. 2, D85748, Garching, Germany}\\
  {\tt $^3$ Dipartimento di Fisica e Astronomia, Alma Mater Studiorum, Universita' di Bologna, Viale Berti Pichat 6, 40127, Bologna, Italy}\\
  {\tt $^4$ INAF-Osservatorio Astronomico di Bologna, Via Ranzani 1, 40127, Bologna, Italy}\\
   {\tt $^5$ Max Planck Institut f\"ur extraterrestrische Physik (MPE),
  Postfach 1312,  D85741, Garching, Germany}\\
  {\tt $^6$ Sorbonne Universite', UPMC Univ Paris 06 and CNRS, UMR 7095, IAP, 98b bd Arago, F-75014, Paris, France}\\
   {\tt $^7$ LAM, Universite' d'Aix-Marseille \& CNRS, UMR7326, 38 rue F.Joliot-Curie, 13388 Marseille Cedex 13, France}\\
 }
 \maketitle
 \begin{abstract}
  We present the clustering properties of a complete sample of 968 radio sources detected at 1.4 GHz by the VLA-COSMOS survey with radio fluxes brighter than 0.15 mJy. 
  92\% have redshift determinations from the Laigle et al. (2016) catalogue.
  Based on their radio-luminosity, these objects have been divided into 644 AGN and 247 star-forming galaxies. 
 By fixing the slope of the auto-correlation function to $\gamma=2$, we find $r_0=11.7^{+1.0}_{-1.1}$ Mpc for the clustering length of the whole sample, while $r_0=11.2^{+2.5}_{-3.3}$ Mpc and $r_0=7.8^{+1.6}_{-2.1}$ Mpc ($r_0=6.8^{+1.4}_{-1.8}$ Mpc for $z\le 0.9$) are respectively obtained for AGN and star-forming galaxies. 
These values correspond to minimum masses for dark matter haloes of $M_{\rm min}=10^{13.6^{+0.3}_{-0.6}}$ $M_\odot$ for radio-selected AGN and 
 $M_{\rm min}=10^{13.1^{+0.4}_{-1.6}}$ $M_\odot$ for radio-emitting star-forming galaxies ($M_{\rm min}=10^{12.7^{+0.7}_{-2.2}}$ $M_\odot$ for $z\le 0.9$). 
 Comparisons with previous works imply an independence of the clustering properties of the AGN population with respect to both radio luminosity and redshift.  
  We also investigate the relationship between dark and luminous matter in both populations. We obtain  $\langle M_*\rangle/M_{\rm halo}\simlt 10^{-2.7}$ for  AGN, and 
$\langle M_*\rangle/M_{\rm halo}\simlt 10^{-2.4}$ in the case of star-forming galaxies. 
Furthermore, if we restrict to $z\simlt 0.9$ star-forming galaxies, we derive  $\langle M_*\rangle/M_{\rm halo}\simlt 10^{-2.1}$,  result which clearly shows the cosmic process of stellar build-up as one moves towards the more local universe.
Comparisons between the observed space density of radio-selected AGN and that of dark matter haloes shows that about one in two haloes is associated with a black hole in its radio-active phase.  This suggests that the radio-active phase is a recurrent phenomenon.
 \end{abstract}

 
\section{Introduction}

The history of large-scale structure studies performed with radio sources dates back to the early '80's 
when the works of Seldner \& Peebles (1981) and Shaver \& Pierre (1989) reported the first detection 
of a slight clustering signal in nearby radio sources. A few years later, Kooiman, Burns \& Klypin (1995) and Loan, Wall \& Lahav
(1997) detected strong anisotropy in the distribution of bright radio
objects from the 4.85 GHz Green Bank and Parkes-MIT-NRAO surveys.
But it was only with the advent of the so-far latest generation of wide area radio surveys
such as FIRST (Faint Images of the Radio Sky at 20 cm; Becker,
White \& Helfand 1995), WENSS (Rengelink et al. 1998) and NVSS (NRAO
VLA Sky Survey; Condon et al. 1998) that high precision clustering measurements were made possible thanks to the 
large number of sources observed by these surveys (Cress et al. 1996;
Rengelink et al. 1998; Magliocchetti et al. 1998; Blake \& Wall 2003;
Overzier et al. 2003; Negrello, Magliocchetti \& De Zotti 2006). 
All the aforementioned works converge at indicating that radio sources are more strongly clustered than optically-selected galaxies.

However, despite the fact that radio sources have the enormous advantage of tracing large-scale structure up to very high ($z\sim 4$) redshifts since their signal does 
not get attenuated by intervening dust, 
they are quite difficult to follow up with optical facilities.
This implies that the overwhelming majority of them will not have estimated redshifts. 
And without redshift information it is not possible to assess the real clustering signal produced by these objects.

A number of works tried to overcome the above issue and used wide-area optical surveys  to provide redshift information at least 
for the more local sources detected in mJy-level radio surveys such as FIRST and NVSS. This allowed to estimate their real-space clustering properties 
up to redshifts $z\simeq 0.5$ (Magliocchetti et al. 2004; Brand et al. 2005; Wake et al. 2008; Fine et al. 2011; Linsday et al. 2014).

In the more recent years, the advent of deep enough, sub-mJy,  radio surveys performed on smaller but well studied fields whereby galaxies are provided with a wealth of multi-wavelength information, has opened a new era for the direct investigation of the environmental and clustering properties of radio sources at all redshifts.  In fact, in this case, most of the sources are endowed with  either photometric or in some cases even spectroscopic redshift determinations. 

The first work that tried to estimate the 3D clustering properties of radio-selected sources on one of such fields is that of Lindsay, Jarvis \& McAlpine (2014b). 
However, due to the combination of the the still relatively high flux limit ($F_{1.4 \rm GHz} >0.09$ mJy) and of the relatively small area covered by the VLA-VIRMOS deep field survey, the Lindsay et al. (2014b) work still does not include enough radio objects to allow for a direct assessment 
of their clustering properties and has to rely on the results obtained from the analysis of the cross-correlation between 1.4 GHz-selected sources  and  
near-infrared galaxies observed on the same field. 

This present work tries to overcome this last limitation and investigates the spatial clustering properties of a complete sample of $F_{1.4 \rm GHz} \ge 0.15$ mJy radio sources taken from the VLA-COSMOS survey (Bondi et al. 2008) by directly assessing their auto-correlation function. Furthermore, since the majority of galaxies  on the COSMOS field are provided with a redshift determination, our analysis will be independent of any assumption on the functional form of $N(z)$. Indeed, the COSMOS field (Scoville et al. 2007) covers a large ($\sim 2$ deg$^2$) area and it is observed with very deep (AB$=25-26$) multi-wavelength data, including imaging in 18 intermediate band filters from Subaru (Taniguchi et al. 2007), which allow to pinpoint emission/absorption lines in the SEDs, and NIR/MIR data from UltraVISTA and IRAC (Splash Survey). The photometry is homogenized and blending is also taken into account, making the quality of the data, the photometric redshifts and the stellar masses (Laigle et al. 2016) among the best available. The availability of reliable photometric redshifts is not limited to normal galaxies but it also assured for the X-ray sources detected by Chandra in a deep and homogeneous manner (Civano et al. 2016, Marchesi et al. 2016).

Perhaps more importantly, the number of radio sources detected with fluxes $F_{1.4 \rm GHz} \ge 0.15$ mJy on the COSMOS field is large enough to allow the entire radio population to be divided into its two major components: radio-active AGN and radio-emitting star-forming galaxies. These two sub-samples are still large enough, and clustering estimates can be provided (although with large uncertainties) independently for both these two classes of sources. This will in turn allow us to draw some conclusions on the relationship between visible and dark matter in the case of both star-forming galaxies and radio-active AGN and also on the life-time of the radio-active AGN phase.

Throughout this paper we assume a $\Lambda$CDM cosmology with $H_0=70 \: \rm km\:s^{-1}\: Mpc^{-1}$ ($h_0=0.7$), $\Omega_0=0.3$,  
$\Omega_\Lambda=0.7$ and $\sigma_8=0.8$.





\section{The Dataset}

\subsection{Radio data}
The VLA-COSMOS Large Project observed the $\sim$2 deg$^2$ of the  COSMOS field at 1.4 GHz with the VLA in the A configuration. This survey,  extensively described in Schinnerer et al. (2004) and Schinnerer et al. (2007), provides continuum radio observations with a resolution of 2$^{\prime\prime}$ and a mean 1$\sigma$ sensitivity of about 10.5 $\mu$Jy in the central 1 deg$^2$ region and of about 15 $\mu$Jy in the outer parts. 

As a matter of fact, as clearly shown in Figs 12 and 13 of Schinnerer et al. (2007), the radio coverage on the COSMOS area greatly varies across the field.  The minimum rms noise level for which the entire field is uniformly sampled is about 0.03 mJy.
Since clustering analyses need high completeness levels, for the purpose of our work we then limited ourselves to consider sources with 1.4 GHz integrated radio fluxes larger than about 5 times the minimum rms level which guarantees a uniform data coverage. 
This corresponds to considering only sources with 1.4 GHz integrated fluxes brighter than 0.15 mJy. The adopted catalogue is then derived from that of Bondi et al. (2008) which, by applying our flux cut, returns 968 radio-selected sources spread all over the COSMOS area. This will be our working sample. 

In order to endow radio sources with a redshift determination, we cross-correlated the above sample with the Laigle et al. (2016) catalogue which provides reliable photometric redshifts ($\sigma_{NMAD} = 0.01$ for galaxies brighter than $I=22.5$) for COSMOS galaxies, with only a handful of outliers. When possible, we used spectroscopic redshifts available within the COSMOS collaboration.
Given the high positional accuracy of both the radio and the optical-near infrared surveys, we fix the matching radius to 1 arcsec. This procedure provides redshift estimates for 891 radio sources, i.e.  $\sim 92$ per cent of the parent sample, with  a negligible fraction (about 0.3\%) expected to be spurious matches. More than half of the the redshifts (508) are spectroscopic. Note that, as shown in Magliocchetti et al. (2014) and (2016b) the above percentage of radio sources with an optical counterpart is roughly independent of radio flux. 
In order to quantify the fraction of radio sources that are also X-ray emitters, we cross matched our catalogue with that of Marchesi et al. (2016) which provides for each source detected in the Chandra Legacy-COSMOS (Civano et al. 2016) the most reliable counterpart and photometric redshift as computed in Salvato et al. (2011). Of the 891 sources, we found that 242 were X-ray detected. Of these, 205 are provided with spectroscopic redshift and  the rest with photometric redshift.
				
\begin{figure*}
\includegraphics[scale=0.27]{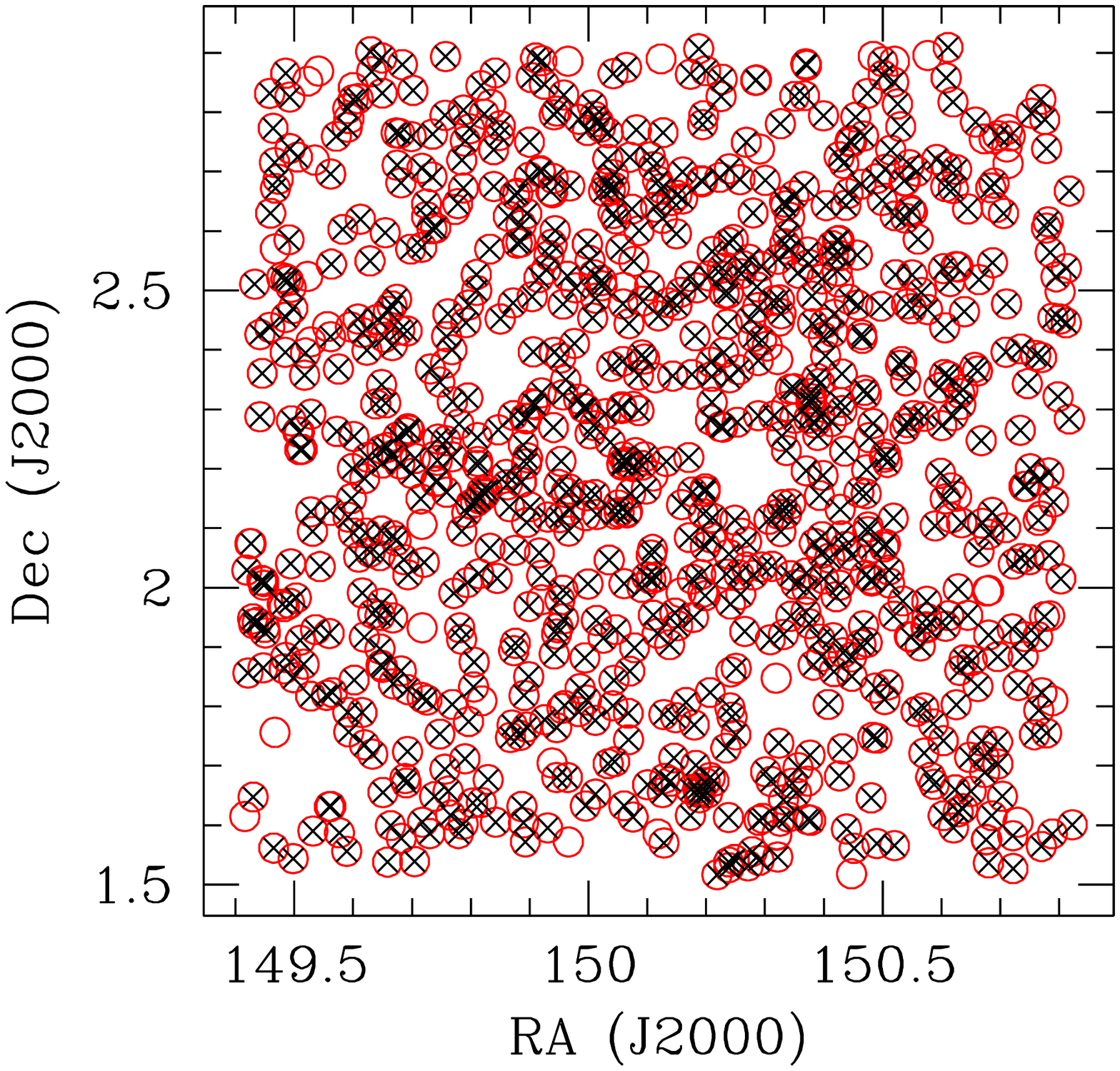}
\includegraphics[scale=0.27]{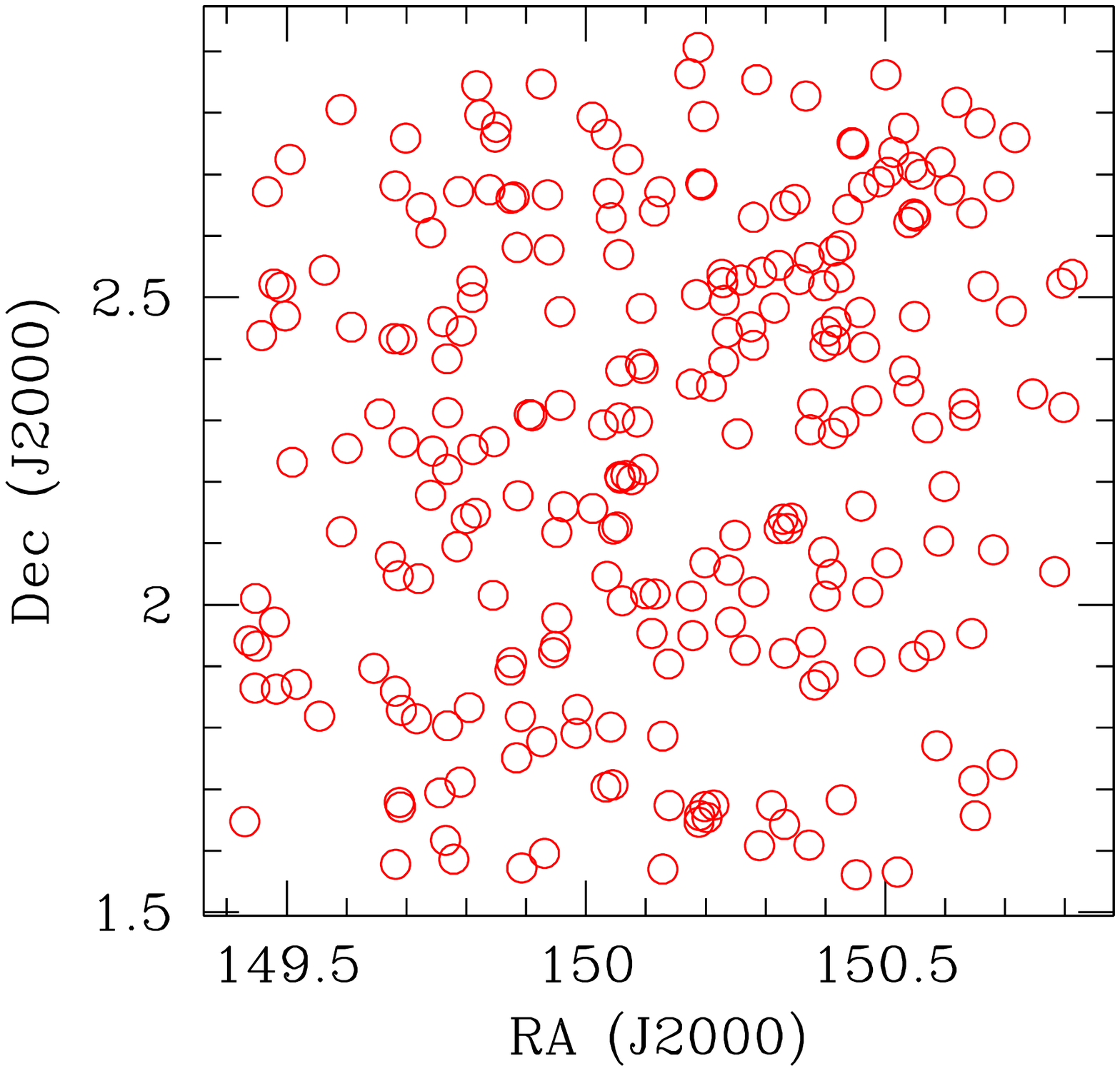}
\includegraphics[scale=0.27]{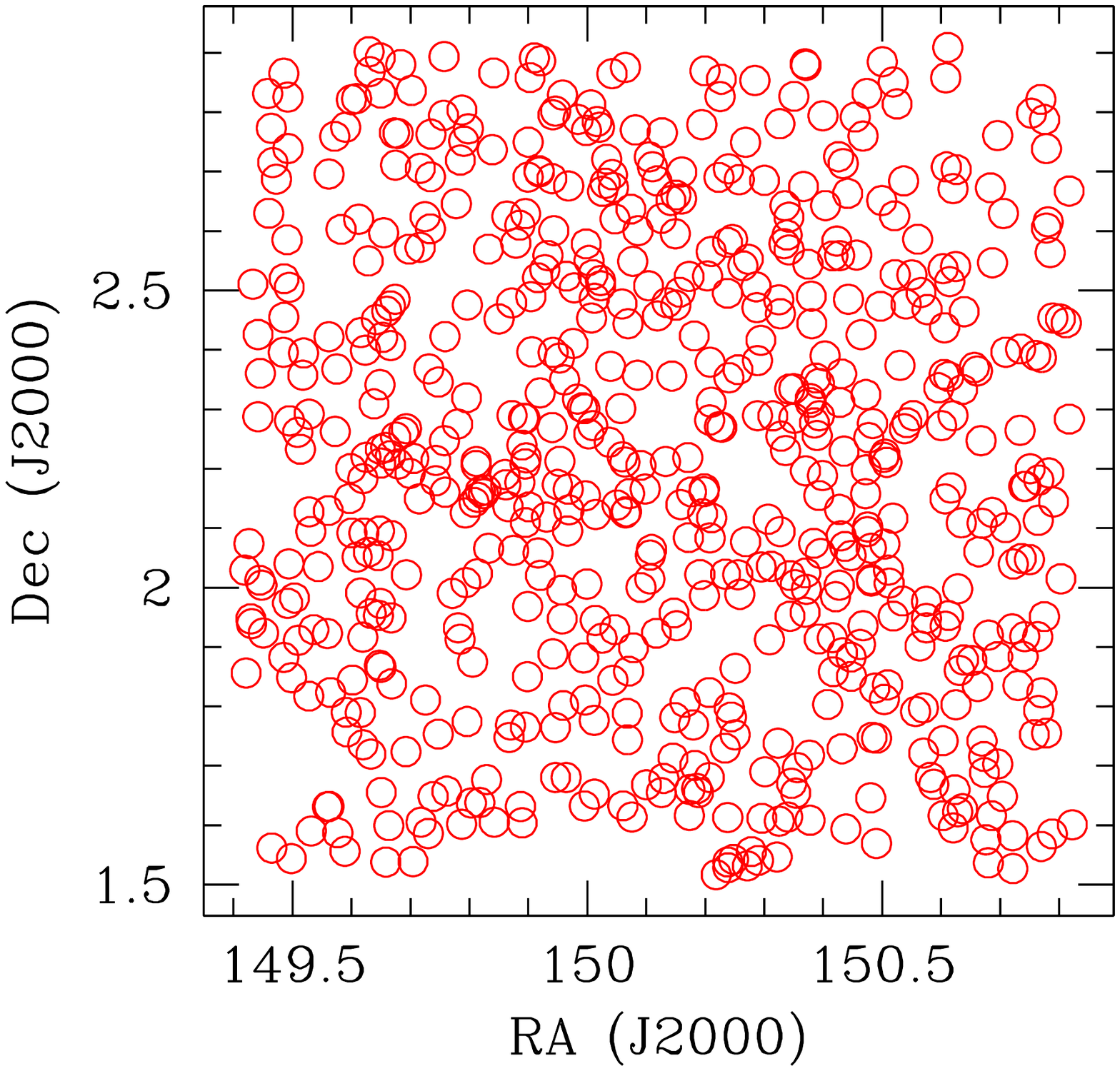}
\caption{Left-hand panel: projected distribution on the COSMOS area of 1.4 GHz-selected sources  with radio fluxes brighter than 0.15 mJy. Open circles represent all sources, while black crosses indicate those which have a redshift determination from the Laigle et al. (2016) catalogue. Middle panel: projected distribution on the COSMOS area of 1.4 GHz-selected star-forming galaxies  with radio fluxes brighter than 0.15 mJy. Right-hand panel: projected distribution on the COSMOS area of 1.4 GHz-selected AGN with radio fluxes brighter than 0.15 mJy.
\label{fig:radecall}}
\end{figure*}

The distribution on the COSMOS area of 1.4 GHz-selected sources brighter than 0.15 mJy is shown in the left-hand panel of Figure \ref{fig:radecall} by the open circles. Sources which also possess a redshift estimate are marked by crosses. 		
The redshift distribution of these sources is presented in Figure \ref{fig:nz} by the solid (black) histogram. It is interesting to note that such a distribution features two major peaks of approximately the same amplitude, one quite local, around $z\sim 0.3$, and another one at $z\sim 1$. Beyond that value the redshift distribution slowly declines, even though sources can be found up to redshifts $z\sim 4$. We will discuss in greater detail about the two peaks in Section 2.2. 
The relevant properties of the sample of radio sources adopted in this work are presented in the first and second rows of Table 1.

\subsection{AGN and star-forming galaxies selection via their radio luminosity}
One of main purposes of this work  is not only to assess the clustering properties of the sub-mJy population of radio sources as a whole, but also to estimate the clustering properties of the two families which contribute to the total radio counts: AGN and star-forming galaxies. 
However, in the absence of spectroscopic information for all the sources in exam (and sometimes even with this information at hand), discerning between AGN-powered and star-formation-powered sources in a monochromatic radio survey might be a tricky issue.

Following the approach introduced by Magliocchetti et al. (2014) and subsequently also used in Magliocchetti et al. (2016; 2016b), we decided to use radio emission as the sole indicator of AGN rather than star forming activity. 
The method bases itself on the  results of McAlpine, Jarvis \& Bonfield (2013) who used the optical and near infrared Spectral Energy Distributions (SED) of a sample of 942  radio sources from the VIDEO-XMM3 field to distinguish between star-forming and AGN-powered galaxies and derive their redshifts and luminosity functions.\\
Investigations of their results show that the radio luminosity $P_{\rm cross}$ beyond which  AGN-powered galaxies become the dominant radio population scales with redshift roughly as
\begin{eqnarray}
\log_{10}P_{\rm cross}(z)=\log_{10}P_{0,{\rm cross}}+z,
\label{eq:P}
\end{eqnarray}
at least up to $z\sim 1.8$. $P_{0,\rm cross}=10^{21.7}$[W Hz$^{-1}$ sr$^{-1}$] is the value which holds in the local universe and which roughly coincides with the break in the radio luminosity function of star-forming galaxies (cfr Magliocchetti et al. 2002; Mauch \& Sadler 2007). Beyond this value,  their  luminosity function steeply declines, and the contribution of star-forming galaxies to the total radio population is drastically reduced to a negligible percentage. The same trend is true at higher redshifts, and since the radio luminosity function of star-forming galaxies drops off in a much steeper way than that of AGN at all $z$, we expect the chances of contamination during the selection process of the two populations to be quite low. 

Radio luminosities for the COSMOS sample of radio-selected sources endowed with a redshift estimate have been calculated according to the relation:
\begin{eqnarray}
P_{1.4 \rm GHz}=F_{1.4 \rm GHz} D^2 (1+z)^{3+\alpha},
\end{eqnarray}
where the result is in [W Hz$^{-1}$ sr$^{-1}$] units, $D$ is the angular diameter distance and $\alpha$ is the spectral index of the radio emission ($F(\nu)\propto \nu^{-\alpha}$). 

As radio sources in the COSMOS region do not have published estimates for the quantity $\alpha$, we then adopted the average value $\alpha=0.7$ found for similar surveys (e.g. Randall et al. 2012 and references therein) both for star-forming galaxies and for AGN emission. Such an assumption is expected to hold since 1) radio sources considered in this work are faint, therefore the chances of finding a large number of bright, flat spectrum AGN are low and 2) recent results report values  $\alpha\simeq 0.7$ also for star-forming galaxies at $z\simeq 2$ (Ibar et al. 2010), similar to what found locally for the same population (Condon 1992). Furthermore, $\alpha=0.7$ is in excellent agreement with the average value obtained by Magliocchetti et al. (2016) in the case of 1.4 GHz-selected sources from  the Lockman Hole ($\langle \alpha\rangle =0.685$, independent of flux and redshift).

We then distinguished between AGN-powered galaxies and star-forming galaxies by means of equation (\ref{eq:P}) for $z\le 1.8$ and by fixing $\log_{10}P_{\rm cross}(z)=23.5$ [W Hz$^{-1}$ sr$^{-1}$ ] at higher redshifts (cfr McAlpine, Jarvis \& Bonfield 2013). 
This procedure identifies 247 star-forming galaxies and 644 AGN. This corresponds to 72\% of the total radio population.
Note that, due to the adopted selection criteria and thanks to the chosen flux limit, the AGN sample is {\it complete} with respect to radio selection up to redshifts $z\sim 2.3$, i.e. below that redshift value, the considered sample  includes  {\it all} radio-emitting AGN selected at 1.4 GHz and endowed with a redshift determination. 

The two distributions of star-forming galaxies and AGN over the COSMOS area are presented in the middle and right-hand panels of Figure \ref{fig:radecall}. Their redshift distributions are instead shown in Figure \ref{fig:nz} respectively by the (blue) dotted line for the population of star-forming galaxies and by the (red) dashed line for AGN. It is interesting to notice that the two-peaked distribution observed for the total, $F_{1.4 \rm GHz}\ge 0.15$ mJy, population is the result of the superposition of these two astrophysical sources, whereby star-forming galaxies are responsible for the peak at $z\simeq 0.3$ and dominate the counts for redshifts below $z\simlt 0.4$, while AGN produce the peak at $z\simeq 1$ and constitute the overwhelming majority of the radio population at all redshifts $z\simgt 0.6$.

As already discussed in the previous paragraphs, since the radio luminosity function of star-forming galaxies at all redshifts is quite steep while that of the radio-selected AGN population is rather flat, we do not expect a large level of contamination in the two considered sub-samples of sources. This obviously does not imply that some contamination will not be possible, especially in the very proximity of the luminosity values where the two luminosity functions cross at the various redshifts. Also, since the two populations have been originally divided by Mc Alpine et al. (2013) on the basis of their optical and near infrared SEDs, it might happen that an AGN detected in the optical/NIR bands is in fact not active at radio wavelengths. The consequence of this effect is that the radio signal observed for that object 
is mistakenly attributed by our method to accretion onto a black hole rather than being correctly identified as originating from star-forming processes within the host galaxy. However, as shown in Magliocchetti et al. (2014), we do not expect many of such cases. Possible effects of the above contamination issues on the clustering results will be tackled in the next Sections.


\section{Clustering properties}

\subsection{The Angular Correlation Function}

The angular two-point correlation function
$w(\theta)$ is estimated by comparing the
distribution of a chosen population of sources with a catalogue of randomly
distributed data subject to the same mask constraints as the
real ones.\\
As for the estimator, we chose to use the one introduced by Hamilton (1993):
\begin{eqnarray}
w(\theta) = 4\times \frac{DD\cdot RR}{(DR)^2} -1, 
\label{eq:xiest}
\end{eqnarray}
where $DD$, $RR$ and $DR$ are the number of data-data, random-random 
and data-random pairs separated by a distance $\theta$. 

Since at the chosen flux level the VLA-COSMOS  dataset is $\ge95$\% complete (cfr Section 2) and the data sampling is largely uniform throughout the field, 
we have estimated $w(\theta)$ by simply generating random catalogues of about 20 times as many sources as the original catalogues which filled the whole surveyed area except for the outermost regions which presented irregular data coverage. 

 $w(\theta)$ in eq. (\ref{eq:xiest})  was then estimated on angular scales ranging from 
$10^{-3}$ degrees to $\theta\sim 0.7$ degrees, 
since the upper limit cannot be larger than about half the maximum scale probed by a survey.

Error-bars have been obtained  from jack-knife resampling. In all three cases, the COSMOS field was divided into 25 quadrants of approximately the same area and the correlation function $w(\theta)$ was calculated for 25 different resampling of the data, 
each one obtained by omitting one quadrant. Errors were then obtained from the variance in $w$. 

The above exercise was repeated three times: one for the whole $F_{1.4 \rm GHz}\ge 0.15$ mJy radio sample, one for the sample of star-forming galaxies and one for the AGN sample. The resulting observed angular correlation functions are shown in Figure \ref{fig:wall}, where the plotted error-bars are the 1$\sigma$ uncertainties obtained via jack-knife resampling as explained above.

If we then assume the standard power-law form for the two-point angular correlation function $w(\theta)=A\theta^{1-\gamma}$,
we can estimate the amplitude $A$ and the slope $\gamma$ by using a least-squares
fit to the data.  The small area of  the COSMOS field introduces a negative bias through the integral constraint $\int w^{\rm est} d \Omega_1 d\Omega_2=0$. We correct for this effect by fitting to $A \theta^{1-\gamma}-AC$, where $C= 0.65$ (for $\gamma=2$) as found by numerical integration following Roche \& Eales (1999). By doing this,  for the whole radio sample we obtain an amplitude $A=2.2^{+3.9}_{-1.5}\cdot 10^{-3}$ and a slope $\gamma=2.0^{+0.2}_{-0.2}$.

Unfortunately, error-bars on the correlation function of star-forming galaxies and AGN are too large to allow for both $A$ and $\gamma$ to be estimated from the data ($\gamma=1.7\pm 0.5$ and $A=[0.5^{+14}_{-0.4}]\cdot 10^{-2}$ for AGN and 
$\gamma=1.9^{+0.5}_{-0.4}$ and $A=[0.7^{+26}_{-0.6}]\cdot 10^{-2}$ for star-forming galaxies). 
In these two latter cases we then decided to fix the value of $\gamma$ to 2, in agreement with that found from the analysis of the clustering properties of the radio population as a whole. 
By doing this, we then obtain: $A=4.3^{+2}_{-2}\cdot 10^{-3}$ in the case of star-forming galaxies and $A=1.6^{+0.8}_{-0.8}\cdot 10^{-3}$ for AGN. We stress that the choice of fixing $\gamma$ to 2, although suitable to the data presented in the middle and left-hand panels of Figure \ref{fig:wall} and in agreement with the best-fitting value obtained for the radio population as a whole, implies that the covariance between the amplitude of the correlation function and its slope has been ignored. This in turn means that the errors associated with $A$ presented in this latter two cases have been underestimated.

As a last point, we note that if we fix the slope $\gamma$ to its best-fit value of 2 also in the process of fitting the observed $w(\theta)$ derived for  the  whole radio population, we obtain an amplitude $A=[2.2\pm 0.4]\cdot 10^{-3}$. All the above values and the associated uncertainties are summarized in Table 1.

\begin{figure}
\includegraphics[scale=0.38]{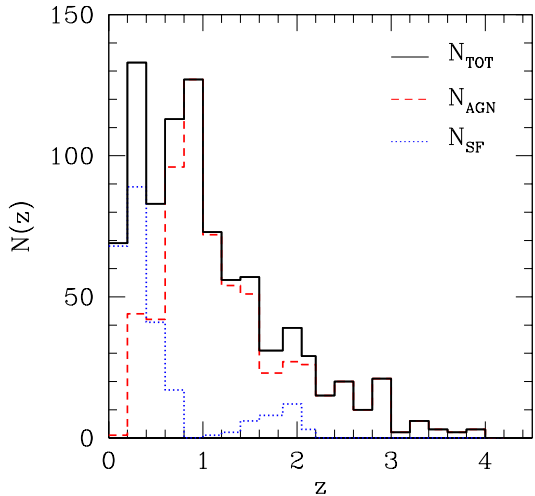}
\caption{Redshift distribution of 1.4 GHz-selected sources with fluxes brighter than 0.15 mJy as found on the COSMOS area. The dashed line indicates AGN, the dotted line star-forming galaxies, while the solid line represents the sum of the two populations.
\label{fig:nz}}
\end{figure}

\subsection{Relation to spatial quantities}

The angular two-point correlation
function $w(\theta)$ is related to the spatial two-point correlation function
$\xi(r,z)$ via the relativistic Limber equation (Peebles,
1980), which requires knowledge of both the cosmological model and of the redshift distribution of the sources under exam.


As already seen in Section 2, COSMOS is provided with a reliable and statistically complete catalogue of source redshifts (either spectroscopic or photometric; Laigle et al. 2016). 
By then assuming a spatial correlation function of the form $\xi(r,z)=(r/r_0)^{-\gamma}$ and by considering the 
redshift distributions of the different radio populations as presented in Figure~\ref{fig:nz}, from the observed angular correlation functions presented in Section 3.1 and for the adopted cosmology, we obtain values for the clustering length $r_0$: $r_0=11.7^{+5.5}_{-3.8}$ Mpc for the whole radio population at a median redshift $\langle z\rangle\sim 1.11$ 
($r_0=11.7^{+1.0}_{-1.1}$ Mpc in case $\gamma$ is fixed to the value of 2), $r_0=7.8^{+1.6}_{-2.1}$ Mpc for star-forming galaxies at a median redshift $\langle z\rangle \sim 0.49$ and $r_0=11.2^{+2.5}_{-3.3}$ Mpc in the case of AGN at a median redshift $\langle z \rangle \sim 1.24$.
All the quoted clustering lengths are comoving.  We remark once again that the choice for a fixed value of the quantity $\gamma=2$ in the case of AGN and star-forming galaxies, implies that the quoted errors on $r_0$ provided for these two populations have been underestimated. All the above quantities are summarized in Table 1.


\begin{figure*}
\includegraphics[scale=0.28]{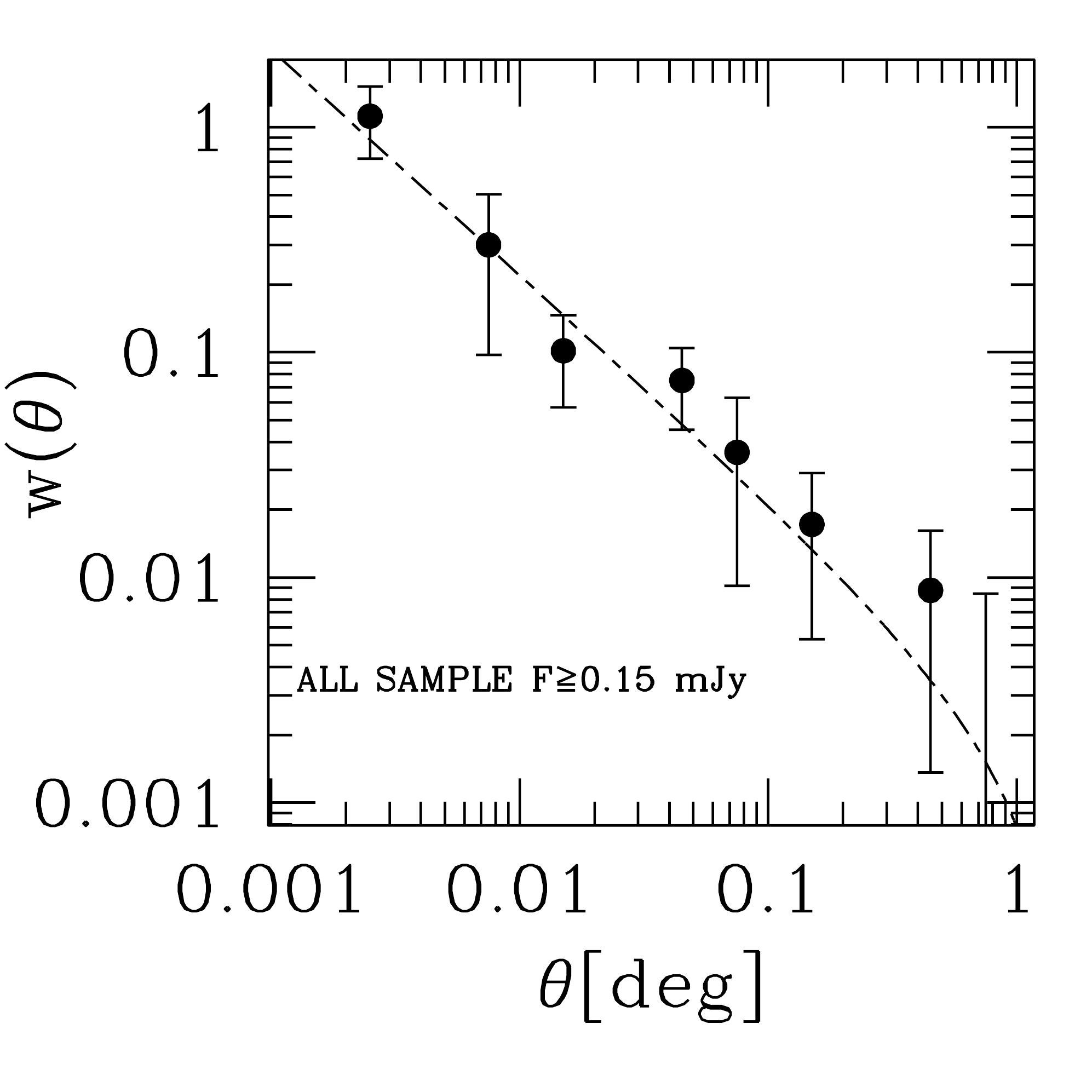}
\includegraphics[scale=0.28]{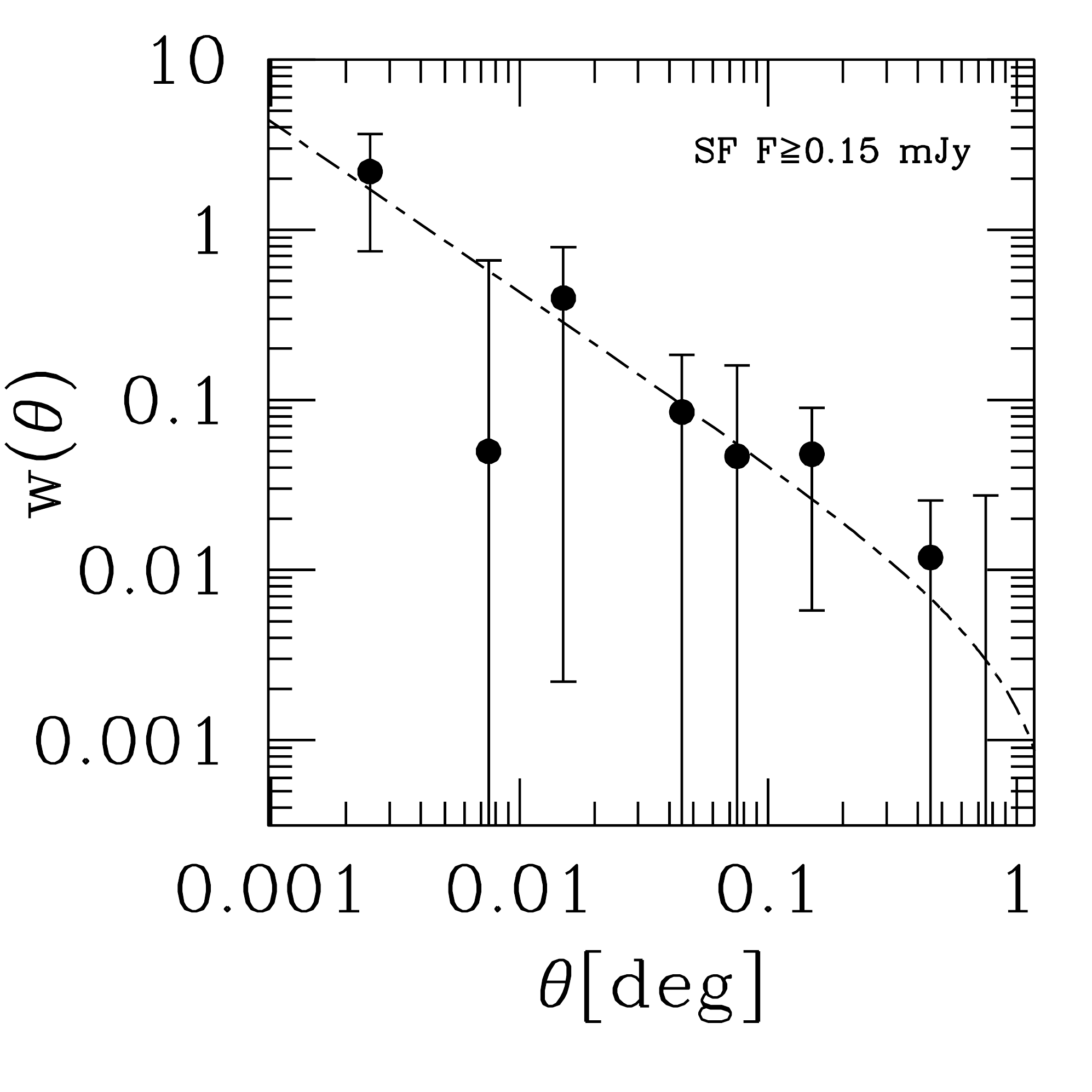}
\includegraphics[scale=0.28]{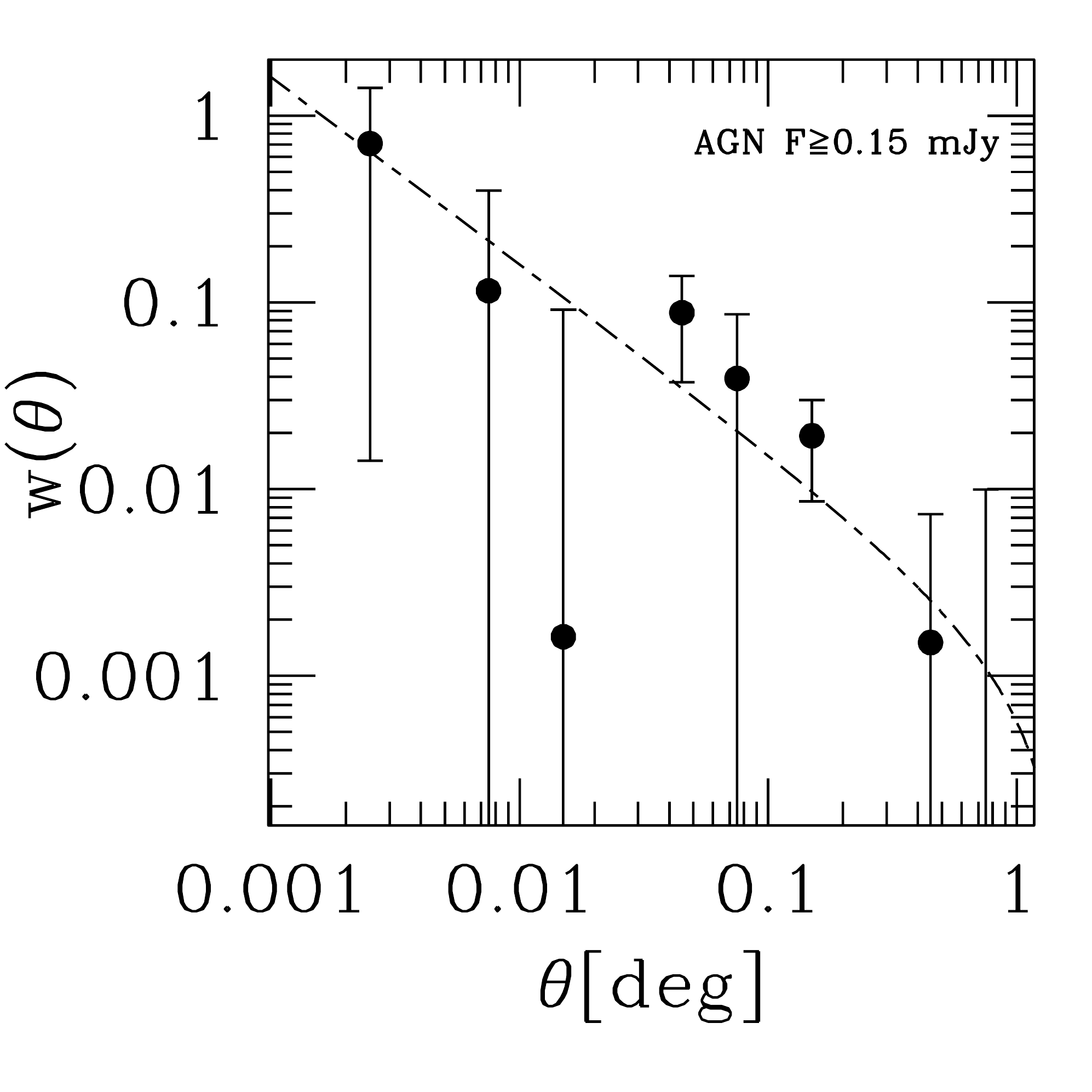}
\caption{Left-hand panel: angular correlation function for all radio sources in the COSMOS-VLA area with $F_{1.4 \rm GHz}\ge 0.15$ mJy. The dashed line indicates the best fit obtained for a functional form $\omega(\theta)=A\theta^{1-\gamma}$, with $A=2.2\cdot 10^{-3}$ and $\gamma=2$. Middle panel: angular correlation function for radio-selected star-forming galaxies (SF) in the COSMOS-VLA area with $F_{1.4 \rm GHz}\ge 0.15$ mJy. The dashed line indicates the best fit obtained for a functional form $\omega(\theta)=A\theta^{1-\gamma}$, with $A=4.3\cdot 10^{-3}$ and $\gamma=2$. Right-hand panel: angular correlation function for radio-selected AGN in the COSMOS-VLA area with $F_{1.4 \rm GHz}\ge 0.15$ mJy. The dashed line indicates the best fit obtained for a functional form $\omega(\theta)=A\theta^{1-\gamma}$, with $A=1.6\cdot 10^{-3}$ and $\gamma=2$.
\label{fig:wall}}
\end{figure*}


\begin{table*}
\begin{center}
\caption{Properties of the sources selected on the COSMOS-VLA area with fluxes $F_{\rm 1.4 GHz}\ge 0.15$ mJy. The first two rows are for the whole sample, while the third one refers to AGN and the fourth and fifth ones to star-forming galaxies.
The first column shows the total number of objects $N$, the second one the number of sources with a redshift determination $N_z$ ($\equiv N$ in the case of AGN and star-forming galaxies), the third one provides the average redshift of the source distribution, the fourth one indicates the best-fit 
values for the amplitude $A$ of the projected correlation function $w(\theta)=A \theta^{1-\gamma}$, the fifth column provides the best-fit value for the slope $\gamma$ (fixed to the value of 2 in the second row and in the case of AGN and star-forming galaxies), the sixth column illustrates the best-fit values for the clustering radius $r_0$ (expressed in [Mpc] units), the seventh column shows the values for the average stellar mass $\langle M_* \rangle$ (expressed in $M_\odot$ units)  for each population, while the last two columns respectively report the best-fit values for the minimum halo mass 
$M_{\rm min}$ (expressed in [$M_\odot]$ units) and 
those for the effective bias, $\langle b_{\rm eff}\rangle$, averaged over the whole redshift range (see text for details).}
\begin{tabular}{llllllllll}
\hline
\hline
&$N$& $N_{\rm z}$&$\langle z\rangle $&$A$& $\gamma$&$r_0$&$\log_{10}$ $\langle M_*\rangle$ &$\log M_{\rm min}$ & $\langle b_{\rm eff}\rangle$\\
\hline
All Sample & 968&891&$1.04\pm 0.76$& $2.2^{+3.9}_{-1.5}\cdot 10^{-3}$&$2.0^{+0.2}_{-0.2}$&$11.7^{+5.5}_{-3.8}$&$10.8\pm 0.5$& $13.8^{+0.2}_{-0.3}$&$4.0^{+0.7}_{-0.8}$\\
All Sample&968&891&$1.04\pm 0.76$&$2.2^{+0.4}_{-0.4}\cdot 10^{-3}$&2.0 (fixed)&$11.7^{+1.0}_{-1.1}$&$10.8\pm 0.5$&$13.8^{+0.2}_{-0.3}$&$4.0^{+0.7}_{-0.8}$\\
AGN&644&644&$1.25\pm 0.74$&$1.6^{+0.8}_{-0.8}\cdot 10^{-3}$ &2.0(fixed)&$11.2^{+2.5}_{-3.3}$&$10.9\pm 0.5$&$13.6^{+0.3}_{-0.6}$&$3.9^{+1.0}_{-1.3}$\\
Star-forming &247&247&$0.50\pm 0.50$&$4.3^{+2.0}_{-2.0}\cdot 10^{-3}$&2.0(fixed)&$7.8^{+1.6}_{-2.1}$&$10.7\pm 0.5$&$13.1^{+0.4}_{-1.6}$&$1.8^{+0.5}_{-0.8}$\\
Star-forming ($z<1$) &215&215&$0.31\pm 0.17$&$4.3^{+2.0}_{-2.0}\cdot 10^{-3}$&2.0(fixed)&$6.8^{+1.4}_{-1.8}$&$10.6\pm 0.5$&$12.7^{+0.7}_{-2.2}$&$1.3^{+0.4}_{-0.7}$\\
\hline
\hline
\end{tabular}
\end{center}
\end{table*}

\begin{figure*}
\includegraphics[scale=0.4]{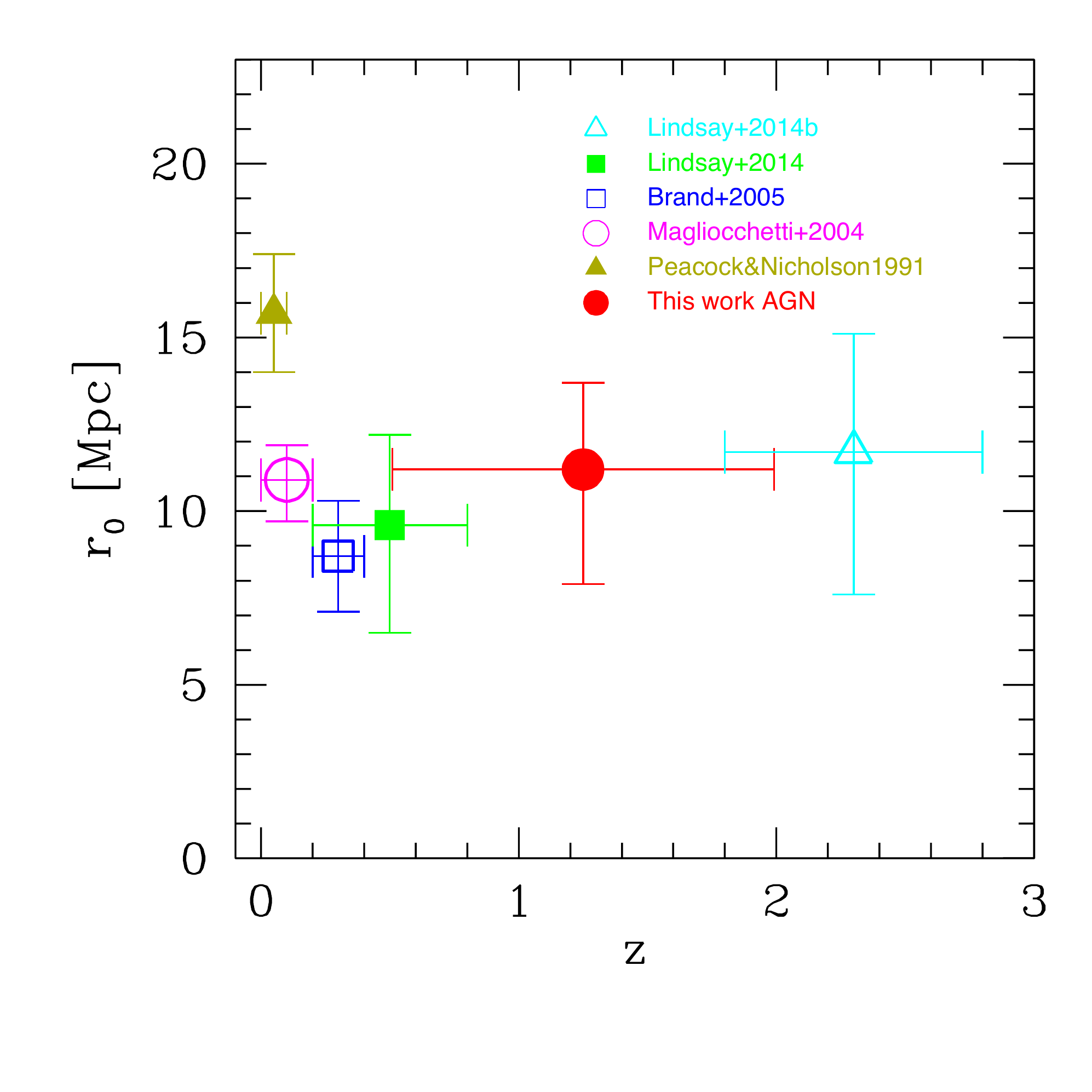}
\includegraphics[scale=0.4]{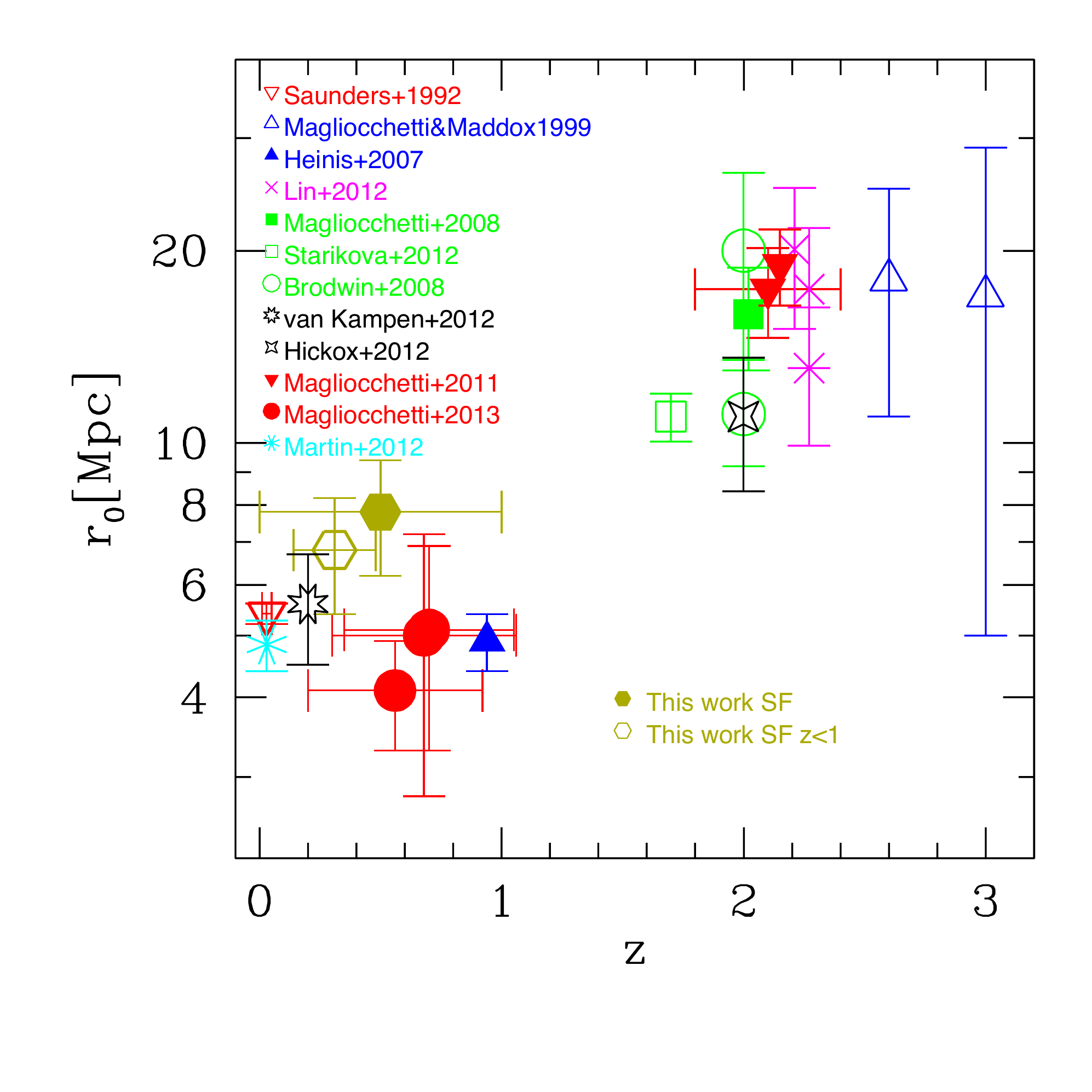}
\caption{Redshift evolution of the clustering length $r_0$ in the case of radio-selected AGN (left-hand panel) and star-forming galaxies (right-hand panel). The left-hand panel compares the results from our work with those of Peacock \& Nicholson (1991), Magliocchetti et al. (2004), Brand et al. (2005), Lindsay et al. (2014) and (2014a). The right-hand panel instead shows the evolution of the clustering length of star-forming galaxies selected in different ways: Far-Infrared (Saunders et al 1992; Magliocchetti et al. 2011; 2013; Van Kampen et al. 2012; Hickox et al. 2012), UV (Magliocchetti \& Maddox 1999; Heinis et al. 2007), Mid-Infrared (Magliocchetti et al. 2008; Brodwin et al. 2008; Starikova et al. 2012), BzK method (Lin et al. 2012), HI emission (Martin et al. 2012) and radio at 1.4 GHz (this work, where the entire population of star-formers is represented by the full hexagon, while the empty one is for more local, $z<1$, sources). Figure adapted from Magliocchetti et al. (2014b).
\label{fig:r0vsz}}
\end{figure*}

We note the value for the clustering length $r_0=11.7^{+5.5}_{-3.8}$  Mpc ($r_0=11.7^{+1.0}_{-1.1}$ Mpc if we fix the slope to $\gamma=2$) and in some cases also for the slope $\gamma\simeq 2$ obtained for the whole population of radio sources agree with those found by Brand et al. (2005) who analyse the clustering properties of a sample of radio objects set at a median redshift $\langle z\rangle \simeq 0.3$, with flux densities $F_{1.4 \rm GHz}\ge 3$ mJy  and optical counterparts coming from the Texas-Oxford NVSS Structure Survey and with those of Lindsay et al. (2014) who instead consider  $\langle z\rangle \simeq 0.5$, $F_{1.4 \rm GHz}\ge 1$ mJy sources from the FIRST (Becker et al. 1995) Survey with optical counterparts in the GAMA (Driver et al. 2011) spectroscopic and photometric maps. 

If we instead concentrate on the results obtained for the population of radio-emitting AGN, we find an excellent agreement between our findings ($r_0=11.2^{+2.5}_{-3.3}$ Mpc) and those of Magliocchetti et al. (2004)  
($r_0\simeq 11$ Mpc) obtained in the case of  AGN optically selected from $F_{1.4 \rm GHz}\ge 1$ mJy FIRST radio sources with a counterpart in the 2dF Galaxy Redshift Survey (Colless et al. 2001) maps in the very local,  $\langle z\rangle \simeq 0.1$, universe. 
Furthermore, they agree with those of Brand et al. (2005) and Lindsay et al. (2014), as the combination of bright radio flux limits and redshift range covered by these latter works in practice implies that the overwhelming majority of the selected sources are indeed radio-active AGN. They are also in agreement with the high-redshift $r_0$ value obtained from Lindsay et al. (2014b) by following their equation (17) under the assumption of $\gamma=2$, consistent with their results,  where, again, the combination of limiting radio flux and redshift range of the sources ensures a minimal contamination of the radio sample due to star-forming galaxies. Instead, they fall slightly short with respect to the value of $r_0\sim  11/ h_0$ Mpc  found in the pioneeristic work of Peacock \& Nicholson (1991) by measuring the redshift-space correlation function of a sample of $z\simlt 0.1$ radio galaxies endowed with radio fluxes $F_{1.4 \rm GHz}> 0.5$~Jy. All the above values for the comoving clustering length of radio-selected AGN are summarized in the left-hand panel of Figure \ref{fig:r0vsz}. What emerges from investigations of the Figure is a substantial independence of the clustering properties of this population with respect to both radio luminosity and redshift. This in turn suggests that radio-active AGN present similar environmental properties at all different radio luminosities and, possibly more importantly, that there has been no evolution in such properties throughout cosmic epochs, from $z\simeq 2.5-3$ down to the very local universe.

On the other hand, our result of $r_0= 7.8^{+1.6}_{-2.1}$ Mpc for the population of radio-selected star-forming galaxies is comparable, although on the slightly high side, with those found in the literature for this class of sources. This is highlighted in the right-hand panel of Figure \ref{fig:r0vsz} (adapted from Magliocchetti et al. 2014b) where we compare the results obtained from our work (shown by the filled, dark gold, hexagon) with those found in the literature for star-forming galaxies selected with different methods. In more details, (blue) triangles are for star-forming galaxies selected in the UV band (Magliocchetti \& Maddox 1999; Heinis et al. 2007), (green) squares and empty circles for star-forming galaxies selected at 24$\mu$m (Magliocchetti et al. 2008; Brodwin et al. 2008; Starikova et al. 2012), filled (red) circles and updown triangles for sources selected in the FIR at $\sim 60 \mu$m (Saunders et al. 1992; Magliocchetti et al. 2011; 2013), (black) empty stars for sources selected in the FIR at $\sim 250 \mu$m (Van Kampen et al. 2012; Hickox et al. 2012),  (magenta) crosses for star-forming galaxies selected with the BzK method (Lin et al. 2012) and finally the (cyan) asterisk is for those sources selected locally because of their HI emission (Martin et al. 2012).

The reason for the slight higher value obtained in our case with respect to those presented in the right-hand panel of Figure \ref{fig:r0vsz} can be attributed to two effects. The first could be found in the contamination of our star-forming sample due to the presence of low-luminosity AGN which, as shown earlier, are much more strongly correlated than star-forming galaxies and consequently present higher correlation lengths. However, as explained in \S 2.2, such a contamination should not be relevant and therefore its effects are not expected to be important to the scopes of our analysis. More important could be the contribution to the total clustering signal due to star-forming galaxies found in the redshift range $z\simeq [1.5-2.2]$ (cfr Figure \ref{fig:nz}). Indeed these sources, independent of the method used to select them, show extremely high correlation lengths (cfr Figure \ref{fig:r0vsz} and the aforementioned literature). If we then remove these sources from our sample, limit our analysis to $z < 0.9$ sources, and re-calculate their clustering properties, what we get is  a value for the clustering length of $r_0=6.8^{+1.4}_{-1.8}$ Mpc (empty, dark gold, hexagon in the right-hand panel of Figure \ref{fig:r0vsz}), which is much more  in agreement with those found for the same class of sources in a similar redshift range.

All the values obtained for our work are summarized in Table 1.


\section{Connection with physical properties: constraints on the halo mass}

The most common way to connect the clustering signal produced by a population of extra-galactic sources with their physical properties is by means of the Halo Bias method (Mo \& White 1996; Sheth \& Tormen 1999). 
However limited by the fact that it only assumes a one-to-one correspondence between the dark matter halo and the extra-galactic source which inhabits it (i.e. excludes multiple occupancy), this method can nevertheless provide a fair description of the data, especially in the case (like ours) in which the clustering signal is measured with a relatively low statistical confidence.   

Briefly, the Halo Bias method writes the spatial two-point correlation function $\xi_{\rm th}(r,z)$ of a chosen population of objects as the product between the two-point correlation function produced by the distribution of the underlying dark matter  $\xi_ {\rm dm}(r,z)$ and the square of the so-called bias function $b_{\rm eff}(M_{\rm min},z)$,  which at a given redshift only depends on the minimum mass $M_{\rm min}$ of the haloes in which the detected sources reside via the relation: 
\begin{equation}
\xi_{\rm th}(r,z)=\xi_ {\rm dm}(r,z) \cdot b_{\rm eff}^2(M_{\rm min},z).
\label{eq:bias}
\end{equation}

The theoretical angular two-point correlation function $w(\theta)_{\rm th}$ predicted by this model 
is then obtained from eq. (\ref{eq:bias}) by projecting it once again by means of  the Limber equation introduced in \S 3.2 and with the redshift distributions $N(z)$'s 
 provided in Figure \ref{fig:nz}. The bias function $b_{\rm eff}(M_{\rm min},z)$ in eq. (\ref{eq:bias}) was calculated by following the prescription of Sheth \& Tormen (1999), while $\xi_ {\rm dm}(r,z)$ -- fully specified for a given cosmological model and  a chosen normalization $\sigma_8$ -- was analytically derived from the approach of the Peacock \& Dodds (1996).

\begin{figure}
\includegraphics[scale=0.4]{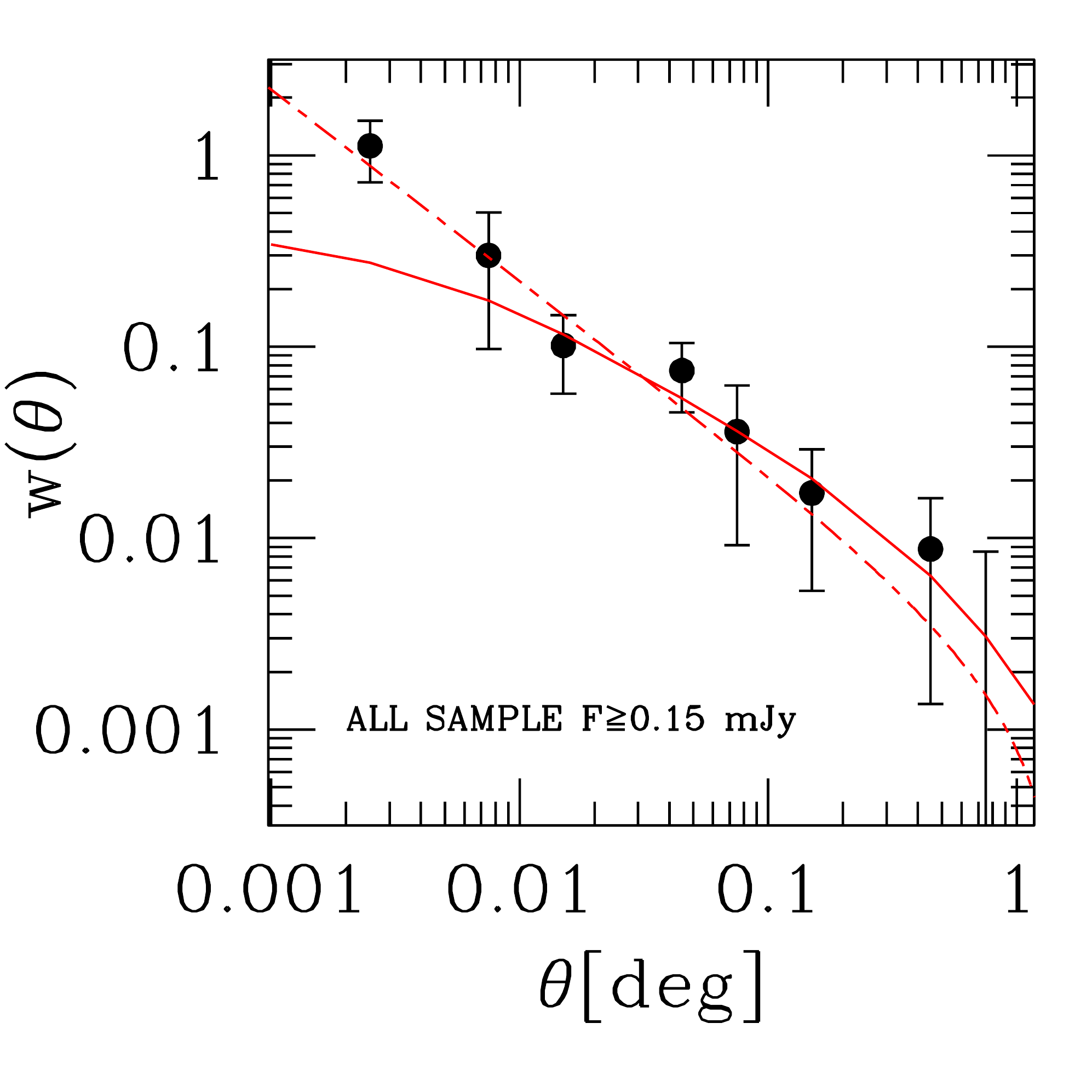}
\caption{Angular correlation function for all radio sources in the COSMOS-VLA area with $F_{1.4 \rm GHz}\ge 0.15$ mJy.  The solid curve is the best-fit model obtained for a minimum halo mass  $M_{\rm min}=10^{13.8}$ $M_\odot$. The dashed line is the same as in Figure  5.
\label{fig:wall_mod}}
\end{figure}	

\begin{figure}
\includegraphics[scale=0.4]{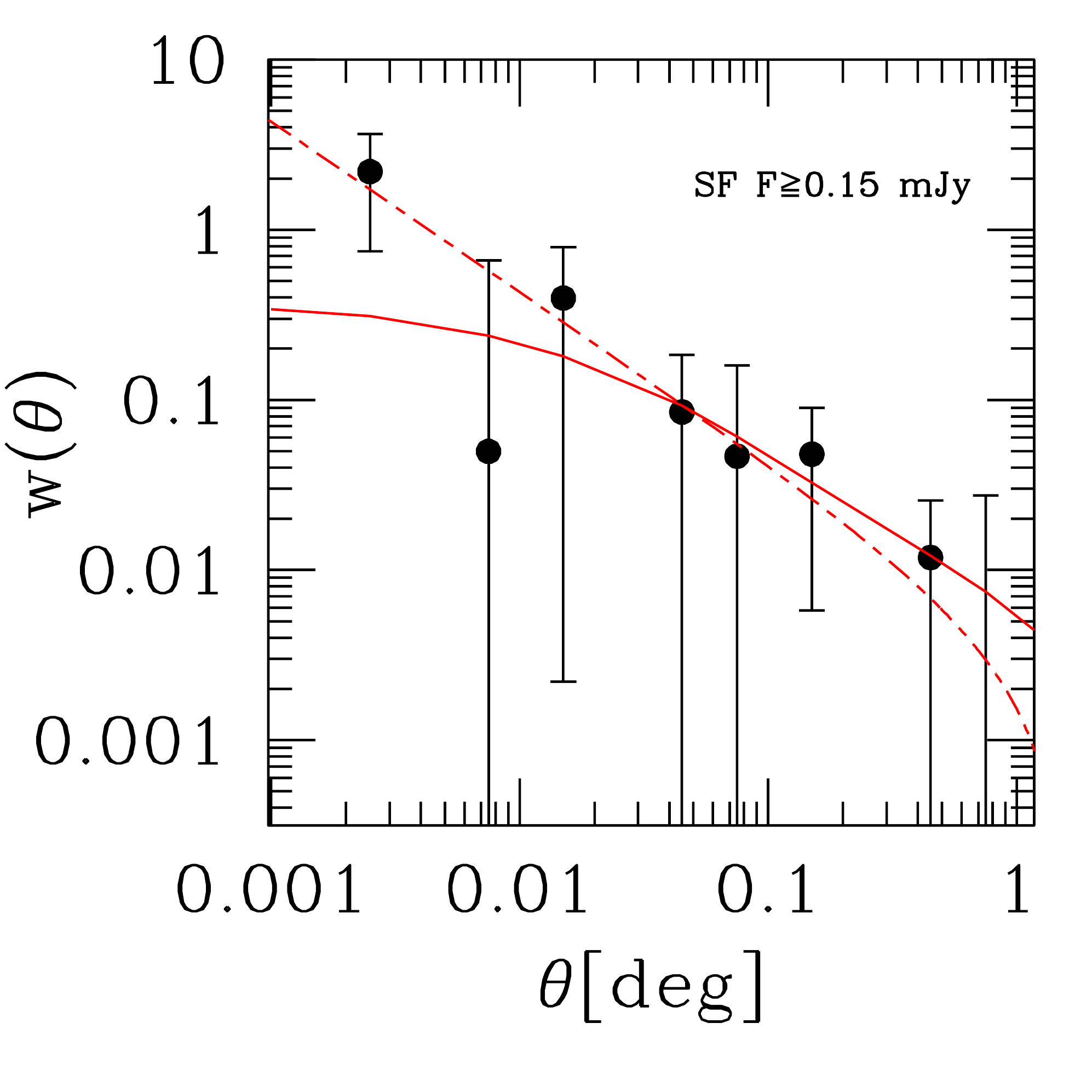}
\caption{Angular correlation function for radio-selected star-forming galaxies in the COSMOS-VLA area with $F_{1.4 \rm GHz}\ge 0.15$ mJy.  The solid curve is the best-fit model obtained for a minimum halo mass  $M_{\rm min}=10^{13.1}$ $M_\odot$. The dashed line is the same as in Figure  6.
\label{fig:wsf_mod}}
\end{figure}	

\begin{figure}
\includegraphics[scale=0.4]{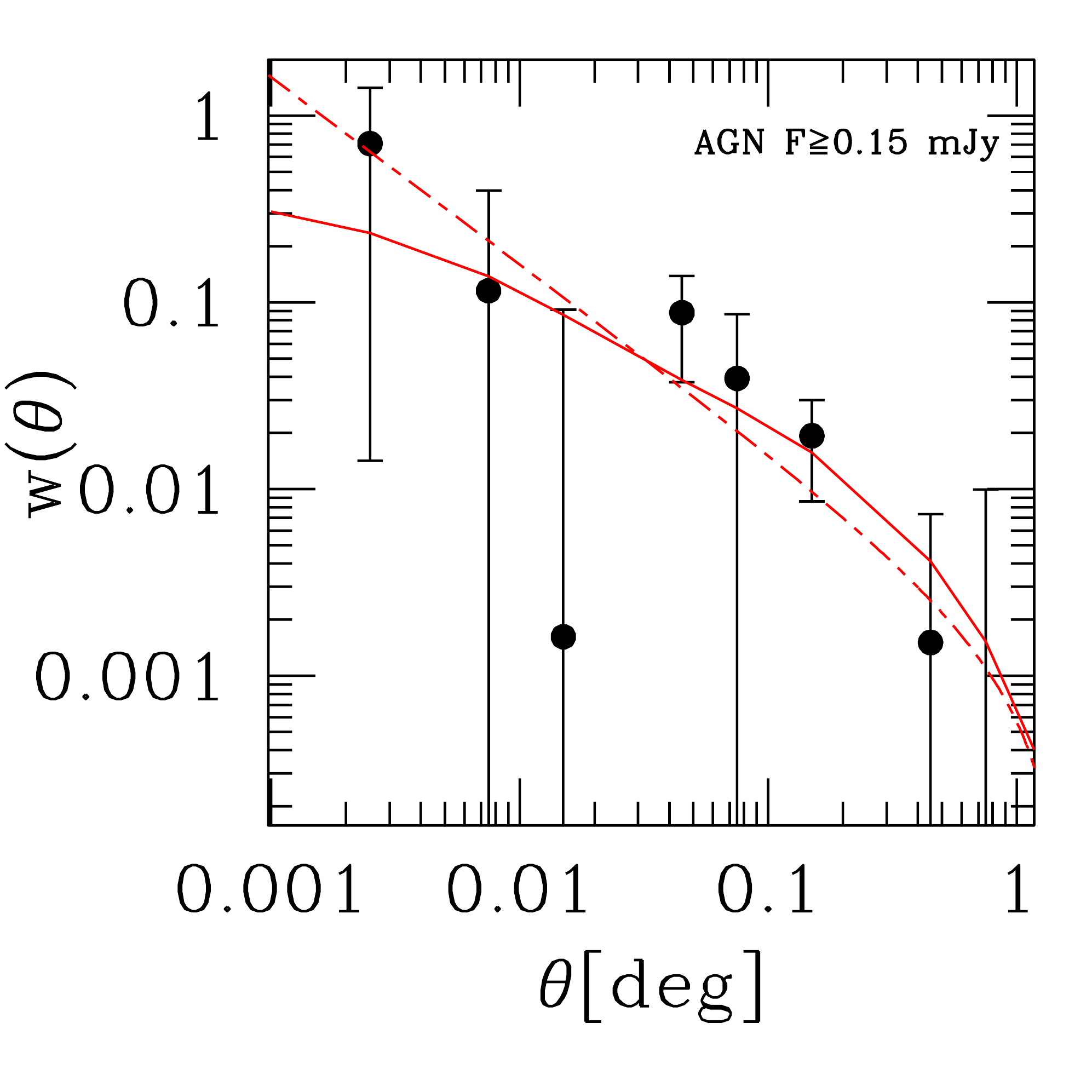}
\caption{Angular correlation function for radio-selected AGN in the COSMOS-VLA area with $F_{1.4 \rm GHz}\ge 0.15$ mJy.  The solid curve is the best-fit model obtained for a minimum halo mass  $M_{\rm min}=10^{13.6}$ $M_\odot$. The dashed line is the same as in Figure  7.
\label{fig:wagn_mod}}
\end{figure}

The resulting angular two-point correlation function $w_{\rm th}$ was then compared to the data, once again by means of a $\chi^2$ fit. As discussed earlier in this Section, it is important to notice that in the present case we are only considering scales outside the galaxy halos. This means that we performed the fit between data and theoretical models only at distances $\simgt 2$ Mpc. At redshifts $\sim 2$ and for the chosen cosmology, this implies considering only angular scales $\theta\simgt 0.01$ degrees. 
 By doing so we obtain our best-fit values for the minimum mass of a halo capable to host the galaxies under exam. In more detail we obtain: $M_{\rm min}=10^{13.8^{+0.2}_{-0.3}}$ $M_\odot$, which corresponds to an effective bias averaged over the whole redshift range redshift  $\langle b_{\rm eff}\rangle=4.0^{+0.7}_{-0.8}$ for the whole radio sample, $M_{\rm min}=10^{13.1^{+0.4}_{-1.6}}$ $M_\odot$  for radio-selected star-forming galaxies ($M_{\rm min}=10^{12.7^{+0.7}_{-2.2}}$ $M_\odot$ if we only consider more local, $z\simlt 0.9$, galaxies)  and $M_{\rm min}=10^{13.6^{+0.3}_{-0.6}}$ $M_\odot$ for AGN. These values and the corresponding results for the effective average bias $\langle b_{\rm eff}\rangle$ are summarized in Table 1. The best-fitting curves are shown in Figures \ref{fig:wall_mod}, \ref{fig:wsf_mod} and \ref{fig:wagn_mod}. Although masked by large uncertainties due to the relative small number of sources in both samples, we note that, as expected, star-forming galaxies are found to reside in halos of  smaller masses than those hosting radio-active AGN. Indeed, this latter population requires masses comparable with those associated with groups or even clusters of galaxies.
 
The values obtained in this work for the halo masses of radio-emitting AGN are in excellent agreement with those found in the literature. For instance,  from investigation of the clustering properties of local, $z\simlt 0.3$,  FIRST-selected radio-active AGN, Magliocchetti et al. (2004) derive halo masses $\simgt 10^{13.4}$ $M_\odot$. The same result is obtained by Hatch et al. (2014) who instead directly investigate the environment of radio-loud sources at redshifts $z > 1.3$ and by  Allison et al. (2015), who measure the bias of FIRST-selected sources at an effective redshift $z\simeq 1.5$ through cross-correlation with lensing. Note that, as already mentioned in \S 3.2, the concordance between different results obtained at different redshifts and for different radio luminosities strongly suggests a lack of cosmological evolution of the clustering properties of radio-selected AGN at all luminosities which, in turn, implies that the environment connected with these sources has not evolved from at least $z\sim 2.5-3$ down to the most local, $z\simlt 0.3$, universe.
   
On the other hand, our results for the population of star-forming galaxies fall high with respect to the values for the halo masses generally found associated with this population of objects (e.g.  Magliocchetti et al. 2013 in the case of FIR-selected galaxies;  Madgwick et al. 2003 or Zehavi et al. 2011 for optically-selected galaxies). As already discussed in \S 3.2, this discrepancy can be explained by either allowing for a non-negliglible contribution of low-luminosity AGN in our sample of radio-emitting star-forming galaxies, or by 
assuming that the high-redshift star-forming galaxies included in our sample boost the value of the measured clustering length.  
As a matter of fact, the aforementioned works consider local sources, while all the most recent Mid-Infrared and Far-Infrared programs aimed at probing star-forming galaxies at high redshifts agree in finding extremely high clustering lengths for this class of objects (e.g. Farrah et al. 2006; Magliocchetti et al. 2007; 2008; Brodwin et al. 2008; Magliocchetti et al. 2011; Starikova et al. 2012. Also cfr the left-hand panel of Figure \ref{fig:r0vsz}). Indeed, if we remove from our working dataset those star-forming galaxies with $z> 0.9$, we end up with a value for the minimum halo mass ($M_{\rm min}=10^{12.7^{+0.7}_{-2.2}}$ $M_\odot$) which is a factor 0.5 dex smaller than that stemming from the analysis of the whole sample. We note that this value is still slightly higher than that expected for this class of sources in the considered redshift range, but within the (large) uncertainties, is nevertheless in agreement with those found in the literature.
   
As a last remark,  it is interesting to notice that, while in the case of star-forming galaxies and of the radio-selected sample as a whole, the best theoretical curves for the projected two-point correlation function $w_{\rm th}(\theta)$ fall short with respect to the data in the small angular regime (cfr Figures \ref{fig:wall_mod} and \ref{fig:wsf_mod}), the same does not happen for the observed correlation function of radio-selected AGN that can be described by the best-fit functional form given in eq. (6) at all scales (cfr Figure \ref{fig:wagn_mod}). This implies that in the case of radio-emitting AGN the assumption of one galaxy per halo made in our analysis is a good one. This should be no surprise: AGN-powered radio galaxies are in fact ubiquitously found at the centres of groups and clusters of galaxies (e.g. Hatch et al. 2014), with a roughly a one-to-one correspondence between radio-AGN and its cluster halo.

\section{The $M_*$/$M_{\rm min}$ relationship in radio-selected AGN and star-forming galaxies and duration of the radio-active phase}

As already mentioned in \S 2, galaxies in the COSMOS field are provided with a large wealth of information on their properties. 
One of them is the stellar mass, $M_*$, which has been derived by Laigle et al. (2016) for the overwhelming majority of such sources.
In our specific case, this is true for 621 radio-selected AGN and 225 star-forming galaxies with 1.4 GHz fluxes brighter than 0.15 mJy. 

With this information in our hand, we can then investigate  the relation between luminous and dark matter in radio-selected AGN and star-forming galaxies by deriving the average stellar mass for each population of  sources and compare 
this value with that of the minimum halo mass of their hosts as determined from clustering results in \S 4.

The average stellar mass for the AGN population is $\langle M_*\rangle=10^{10.9\pm 0.5}$ $M_\odot$, while that for the population of star-forming galaxies is  $\langle M_*\rangle=10^{10.7\pm 0.5}$ $M_\odot$ (cfr Table 1). The two values are indistinguishable within the errors, even though the star-forming population seems to show a mild preference for lower stellar masses. By making use of the results derived in \S 4, we can then estimate the ratio between visible and dark matter in these two cases. We obtain $\langle M_*\rangle/M_{\rm halo}\simlt 10^{-2.7}$ in the case of radio-detected AGN, and 
$\langle M_*\rangle/M_{\rm halo}\simlt 10^{-2.4}$ in the case of radio-emitting star-forming galaxies. Although affected by large uncertainties, a comparison between these results seems to indicate a larger relative stellar content in galaxies which are undergoing a process of global star-formation with respect to those which host a radio-active AGN. The situation becomes somehow more clear if we only concentrate on more local, $z\simlt 0.9$, star-forming galaxies. In fact, in this latter case we obtain: $\langle M_*\rangle=10^{10.6\pm0.5}$ $M_\odot$, and $\langle M_*\rangle/M_{\rm halo}\simlt 10^{-2.1}$, result which shows the cosmic process of star-formation build-up, and indicates a  larger and larger relative stellar content in galaxies identified as star-formers as we approach the more local universe. 

More relevant information on the sources considered in this work come from a direct comparison between their observed space density and that expected in a $\Lambda$CDM universe. 
We stress that this kind of analysis is only possible in the case of radio-selected AGN which, as extensively explained in \S 2, due to the combined effect of the depth of radio observations and of the almost 100\% completeness of the photometric/spectroscopic dataset  on the COSMOS area, constitute a complete sample up to redshifts $z\simeq 2.3$. 

The number of radio-selected AGN on the COSMOS area with photometric or spectroscopic redshifts, $z\le2.3$, is 571. Their space density is $\rho_{\rm AGN}^{\rm obs}(z\le 2.3)=[3.1\pm 0.1]\times 10^{-6}$ Mpc$^{-3}$. On the other hand, the space density of dark matter haloes with masses larger than $10^{13.6}$ M$_\odot$ (cfr \S 4) and redshifts $z\le 2.3$ can be easily obtained via straightforward integration of the Sheth \& Tormen (1999) mass function. By doing this, we obtain $\rho_{\rm AGN}^{\rm th}(z\le 2.3)=7.7 \times 10^{-6}$ Mpc$^{-3}$. 
The ratio between these two quantities is $\sim 0.4$, and corresponds to the fraction of dark matter haloes of masses larger than the value obtained via clustering analysis which is observed to host a radio-active AGN during the time-span which goes from  $z=2.3$ to the local universe. 
This number is rather large and implies that about one in two haloes more massive than $10^{13.6}$ $M_\odot$ is associated with a black hole in its radio-active phase. We note that this result is in full agreement with those of e.g. Hatch et al. (2014), who however, base their conclusions on less solid grounds.

 If we assume that every halo more massive than the above value of $10^{13.6}$ $M_\odot$ hosts a black hole which at some point between $z=2.3$ and $z=0$ will eventually become radio-active, we derive for the life-time of the radio-active phase $\tau\sim 1$ Gyr.  This number is however much larger than the value of a few$\times 10$ Myr obtained for the radio-bright phase of a radio-loud AGN (Blundell \& Rawlings 1999), so our data indicate that the radio-active phase is a recurrent phenomenon, whereby each host halo undergoes multiple radio-active episodes between $z=2.3$ and $z=0$.

\section{CONCLUSIONS}

By making use of deep radio data coming from the VLA-COSMOS survey and of the exquisite catalogue of spectroscopic and photometric redshifts by 
Laigle et al. (2016) provided for galaxies on the COSMOS field, we have identified 968 radio sources down to a 1.4 GHz flux limit of 0.15 mJy. 
891 of such sources are also endowed with a redshift determination, which spans from $z\simeq 0$ to $z\simeq 4$.

These objects have then been divided into two distinct populations: those where the radio signal stems from star-forming activity, and those where radio emission originates from 
AGN activity. The distinction was made only on the basis of the radio luminosity of the considered sources and returns 644 radio-active AGN and 247 radio-emitting star-forming galaxies.

The clustering properties of these objects have then been estimated by means of the projected two-point correlation function $w(\theta)$, which was subsequently deprojected by making use of the 
observed redshift distribution of the considered sources, in order to determine the strength of their clustering via the comoving correlation length $r_0$.
By doing so, for a fixed value of the slope $\gamma=2$ in the expression for the spatial two-point correlation $\xi(r)=\left(r/r_0\right )^{- \gamma}$, we obtain $r_0=11.7^{+1.0}_{-1.1}$ Mpc for the whole sample of radio-selected 
sources. $r_0=11.2^{+2.5}_{-3.3}$ Mpc and $r_0=7.8^{+1.6}_{-2.1}$ Mpc are instead derived respectively for radio-active AGN and star-forming galaxies. 

 These values for the clustering length correspond to minimum masses for dark matter haloes capable to host at least one of such galaxies of $M_{\rm min}=10^{13.6^{+0.3}_{-0.6}}$ $M_\odot$ for radio-selected AGN and 
 $M_{\rm min}=10^{13.1^{+0.4}_{-1.6}}$ $M_\odot$ for radio-emitting star-forming galaxies at all redshifts. 
 
 The values obtained both for the clustering length $r_0$ and for the minimum halo mass $M_{\rm min}$ in the case of radio-active AGN are in excellent agreement with those found in most of the literature (e.g. Magliocchetti et al. 2004; Brand et al. 2005; Linsday et al. 2014; Hatch et al. 2014: Allison et al. 2015 just to mention a few). This consistency amongst different results tends to imply an independence of the clustering properties of such a population with respect to both radio luminosity and redshift, i.e.  similar environmental properties at all different radio luminosities and, possibly more importantly, no evolution in such properties throughout cosmic epochs, from $z\simeq 2.5-3$ down to the very local universe.
 However, they fall short with respect to the results from the works by Wake et al. (2008) and Fine et al. (2011). Indeed for the clustering length associated to their samples the first authors find values which range between $r_0\simeq 7.6 $ $h_0^{-1}$ Mpc and $r_0\simeq 9.5$ $h_0^{-1}$ Mpc at redshifts $z\sim 0.2$, while for $z\sim 0.55$ they obtain $r_0$ in the range $[8.3-9.6]$ $h_0^{-1}$ Mpc. The second authors instead derive $r_0\simeq 9.5$  $h_0^{-1}$ Mpc, $r_0\simeq 9.1$ $h_0^{-1}$ Mpc, $r_0\simeq 8.7$ $h_0^{-1}$ Mpc respectively at $z\sim 0.3$, $z\sim 0.5$ and $z\sim 0.7$. This is probably due to the fact that both these works use data obtained for luminous red galaxies (LRG) in order to identify their radio sources. As it is well known that the LRG population is mainly made of very massive galaxies, it is most likely that a cross-match between radio objects and LRGs preferentially returned optical information for only the most massive objects, therefore leading to a bias of  the clustering results towards high values of both the clustering length and of the minimum halo mass. However, it is interesting to note that within the associated uncertainties even in this case no evolution of the clustering length with cosmic epoch was found.

On the other hand, the clustering results holding for the star-forming galaxy population seem to be higher than those generally found for this class of sources (e.g. Saunders et al. 1992;  Madgwick et al. 2003; Zehavi et al. 2011: Magliocchetti et al. 2013 just to mention a few.  This can be due to two factors: 1) a contribution from low-luminosity AGN which contaminate the star-forming sample and 2) the presence within the star-forming sample of high-redshift galaxies. Both these classes of sources are shown to be very highly clustered and associated to dense environments (e.g. Farrah et al. 2006; Magliocchetti et al. 2008; Brodwin et al. 2008; Magliocchetti et al. 2011; Starikova et al. 2012), so that the presence of either one (or both) of them would result in a boosted clustering signal. 
The second explanation seems more likely, as chances for contamination between the two populations of radio-selected sources are expected to be quite low (cfr Magliocchetti et al. 2014). If we instead restrict the analysis to low-redshift, $z\le 0.9$,  star-forming sources we obtain 
$r_0=6.8^{+1.4}_{-1.8}$ Mpc and $M_{\rm min}=10^{12.7^{+0.7}_{-2.2}}$ $M_\odot$ respectively for the clustering length and for the minimum halo mass, values which are in much better agreement with those found in the literature for the same class of sources within a comparable redshift range. 

As a further step, we used information for galaxies in the COSMOS field  to compute the average stellar mass of both radio-selected AGN and for the star-forming population. Comparisons with the values for the minimum halo mass returned from clustering studies allow to investigate the relationship between dark and luminous matter in both populations. We obtain  $\langle M_*\rangle/M_{\rm halo}\simlt 10^{-2.7}$ for radio-detected AGN, and 
$\langle M_*\rangle/M_{\rm halo}\simlt 10^{-2.4}$ in the case of radio-emitting star-forming galaxies, results which seem to indicate a larger relative stellar content in galaxies which are undergoing a process of global star-formation with respect to those which host a radio-active AGN. Furthermore, if we restrict our attention on more local, $z\simlt 0.9$, star-forming galaxies, we derive  $\langle M_*\rangle/M_{\rm halo}\simlt 10^{-2.1}$, finding which shows the cosmic process of star-formation build-up as one moves towards the more local universe.

Lastly, by comparing the observed space density of radio-selected AGN on the COSMOS-VLA field with that of dark matter haloes more massive than $M_{\rm min}=10^{13.6}$ $M_\odot$ expected from theoretical calculations in a $\Lambda$CDM universe, 
we find a ratio between these two quantities $\sim 0.4$. This result  implies that about one in two haloes more massive than the above value is associated with a black hole in its radio-active phase. 
If we then assume that each one of such haloes hosts a black hole which at some point  will become radio-active, we derive for the life-time of the radio-active phase, $\tau\sim 1$ Gyr.  This number is however much larger than the value of a few$\times 10$ Myr obtained for the radio-bright phase of a radio-loud AGN (Blundell \& Rawlings 1999), so our data indicate that the radio-active phase is a recurrent phenomenon, whereby each host halo undergoes multiple radio-active episodes between $z=2.3$ and $z=0$.\\
\\
\noindent {\bf Acknowledgements}
MM and MB wish to thank  the DFG cluster of excellence 
'Origin and Structure of the Universe'
 (www.universe-cluster.de) for partial support during the completion of this work. 
 We also wish to thank the anonymous referee for constructive comments.

\end{document}